\documentclass[11pt,epsf,letterpaper]{article}
\usepackage{setspace}
\usepackage{color}
\usepackage{amsmath}
\usepackage{amsfonts}
\usepackage{verbatim}
\usepackage{amssymb}
\usepackage{cite}
\usepackage{graphicx}
\usepackage{subcaption}
\usepackage{epstopdf}
\usepackage{epsf}
\usepackage[utf8]{inputenc}
\usepackage[T1]{fontenc}

\definecolor{darkgreen}{rgb}{0,0.35,0}

\providecommand{\U}[1]{\protect\rule{.1in}{.1in}}
\def\ut#1{\rlap{\lower1ex\hbox{$\sim$}}#1{}}

\newcommand{\be}{\nopagebreak[3]\begin{equation}}
\newcommand{\ee}{\end{equation}}
\newcommand{\bee}{\nopagebreak[3]\begin{equation*}}
\newcommand{\eee}{\end{equation*}}
\newcommand{\ba}{\nopagebreak[3]\begin{eqnarray}}
\newcommand{\ea}{\end{eqnarray}}
\newcommand{\baa}{\nopagebreak[3]\begin{eqnarray*}}
\newcommand{\eaa}{\end{eqnarray*}}

\DeclareFontFamily{U}{rsfs}{}         
\DeclareFontShape{U}{rsfs}{m}{n}{<5> rsfs5 <6><7> rsfs7            <8><9><10><10.95><12><14.4><17.28><20.74><24.88> rsfs10}{}     
\DeclareMathAlphabet{\mathfs}{U}{rsfs}{m}{n}

\newcommand{\bseq}{\nopagebreak[3]\begin{subequations}}
\newcommand{\eseq}{\end{subequations}\noindent}

\def\i{i}
\def\o{o}
\def\pb#1{\rlap{\lower1.5ex\hbox{$\longleftarrow$}}{#1}}
\def\dpb#1{\rlap{\lower1.5ex\hbox{$\Longleftarrow$}}{#1}}
\def\spb#1{\rlap{\lower1.5ex\hbox{$\leftharpoonup$}}{#1}}
\def\sdpb#1{\rlap{\lower1.5ex\hbox{$\Leftarrow$}}{#1}}

\def\be{\begin{equation}}
\def\ee{\end{equation}}
\def\ba{\begin{eqnarray}}
\def\ea{\end{eqnarray}}

\input{tcilatex}
\begin{document}

\title{Exact \textit{pp}-waves, (A)dS waves and Kundt spaces in the Abelian-Higgs model}
\author{Fabrizio Canfora$^{1}$, Adolfo Cisterna$^{2}$, Diego Hidalgo$^{1,3,4}$%
, Julio Oliva$^{3}$ \\
$^{1}$\textit{Centro de Estudios Cient\'ificos (CECs), Casilla 1469,
Valdivia, Chile.}\\
$^{2}$\textit{Sede  Esmeralda,  Universidad  de  Tarapacá, Av.   Luis  Emilio  Recabarren  2477,  Iquique,  Chile.}\\
$^{3}$\textit{Departamento de F\'isica, Universidad de Concepci\'on, Casilla
160-C, Concepci\'on, Chile.}\\
$^{4}$ \textit{Instituto de Ciencias F\'isicas y Matem\'aticas, Universidad
Austral de Chile, Valdivia, Chile.}\\
{\small canfora@cecs.cl, cisternamaster@gmail.com, dhidalgo@cecs.cl,
juoliva@udec.cl }}
\maketitle

\begin{abstract}
We find new exact solutions of the Abelian-Higgs model coupled to General Relativity, characterized by a non-vanishing superconducting current. The solutions correspond to \textit{pp}-waves, AdS waves, and Kundt spaces, for which both the Maxwell field and the gradient of the phase of the scalar are aligned with the null direction defining these spaces. In the Kundt family, the geometry of the two-dimensional surfaces orthogonal to the superconducting current is determined by the solutions of the two-dimensional Liouville equation, and in consequence, these surfaces are of constant curvature, as it occurs in a vacuum. The solution to the Liouville equation also acts as a potential for the Maxwell field, which we integrate into a closed-form. Using these results, we show that the combined effects of the gravitational and scalar interactions can confine the electromagnetic field within a bounded region in the surfaces transverse to the current. 
\end{abstract}

\section{Introduction}

A very important step towards a deep understanding of a classical field theory is a proper understanding of its classical solutions. For a generic field theory, this may seem an insurmountable task since the space of solutions are infinite-dimensional, nevertheless, for General Relativity (GR), important classification schemes are available which allow defining classes of solutions, contributing to the understanding of their potential realization in nature \cite{Griffiths:2009dfa}. 

One of the most relevant field theory (both at the classical and quantum level)
is the Abelian-Higgs model (the Maxwell-Ginzburg-Landau theory) which can describe successfully many important semi-classical features of superconductors (see \cite{1}\ and 
\cite{bookVS}\ for detailed reviews: in the following, we will consider the
relativistic version of the theory). A further important phenomenological
implication of this theory is the presence of vortices discovered by
Abrikosov, Nielsen and Olesen in \cite{abrikosov,Nie-Ole}. These are
some of the many reasons why the minimal coupling of the
Abelian-Higgs model with GR has been deeply
investigated (see \cite{bookVS}\ and references therein). Moreover, a
no-hair theorem was proved in \cite{AyonBeato:1996kf}, which can be circumvented for
horizons pierced by a vortex both in the static case \cite{Achucarro:1995nu}, as
well as for stationary black holes \cite{Ghezelbash:2001pq}, and for planar AdS
black holes \cite{Dehghani:2002qc}. In the holographic setup, this system allows constructing holographic superconductors, where near the horizon of a black
hole the scalar acquires a vev \cite{Gubser:2008px,Hartnoll:2008vx,Horowitz:2008bn,Arean:2010zw},
which can be understood as arising due to an instability triggered by a
violation of the effective Breitenlohner-Freedman bound \cite{Breitenlohner:1982bm} in the AdS$_{2}$ near horizon geometry of extremal Reissner-Nordstr\"{o}m black hole. Finally,
this system also finds applications in the holographic description of
superfluidity (see e.g. \cite{Brihaye:2010ce,Arean:2010zw,Brihaye:2011vk}). Given the relevance
of this field theory, it is of uttermost importance to continue shedding
light on the structure of its space of solutions. This paper is devoted to
such a task. In the present manuscript, we will study, with analytic
methods, the gravitational consequences of the presence of a superconducting
current in the Einstein-Maxwell-Ginzburg-Landau theory. 

Of course, one may wonder why to insist on finding analytic solutions if these equations can be solved numerically. Indeed, numerical techniques were already available in the literature
of the eighties and nineties to
analyze these configurations in the gravitating Abelian-Higgs model (see \cite{bookVS}\ and references therein). Despite this, there
are indisputable arguments that strongly suggest that, whenever it is
possible, we should strive for analytic solutions. For example, much of what we currently know about black hole physics in GR, and instantons and monopoles in gauge theories arose from a careful study of the available analytic solutions like the Kerr solution in the former and non-Abelian monopoles and instantons in the latter. Consequently, an analytic tool to
analyze the gravitational effects of superconducting currents in the model relevant to our present study can greatly enlarge our understanding
of this system. Secondly and more concretely, our analysis discloses a nice
mechanism that, at least in principle, can confine the electromagnetic field in the two-dimensional surfaces orthogonal to the superconducting currents.\ 

One may think that pursuing an analytic approach in this non-linear system is hopeless. Nevertheless, the methods
developed in \cite{crystal1,gaugsk,Canfora:2018clt,
lastEPJC,crystal2,crystal3} to propose a proper ansatz, allowed to
construct analytic gauged solitons in the gauged Skyrme model thanks to a
suitable choice of variables which enables to partially decouple the field equations. These were extended in \cite{firstube} to include the minimal coupling with GR and here we show that they are suitable to analyze Einstein-Maxwell-Ginzburg-Landau theory, as well.

In Section II, we present the model and describe properties of the superconducting current supporting the solutions of the following sections. In Section III, we construct the \textit{pp}-wave as well as the AdS-wave solutions and for the latter, in a particular case, we can integrate the whole system in an explicit, closed manner. Section IV is devoted to the construction of the Kundt solutions, characterized by the existence of a null, geodesic, congruence that is not covariantly constant but has vanishing optical scalars. Liouville equation naturally emerges in the constant $u,v$ sector and we obtain non-trivial solutions for both, the positive and negative cosmological constant value. Finally, we provide some conclusions in Section V.

\section{The model}

The gravitating Abelian-Higgs model is described by the action 
\begin{equation}
S[g,\Psi ,A]=\int d^{4}x\sqrt{-g}\left( R-2\Lambda-\frac{1}{4}F_{\mu \nu
}F^{\mu \nu }-D_{\mu }\Psi \,(D^{\mu }\Psi )^{\star }-V(\Psi )\right) ,
\label{actionmodel}
\end{equation}%
where $g$ is the determinant of the metric, $R$ is the Ricci tensor scalar, $%
\Lambda $ the cosmological constant, and we have set $16\pi G=1$. The scalar
field $\Psi $ is complex, and $\Psi ^{\star }$ denotes its complex
conjugate. The electromagnetic field strength is given by $F_{\mu \nu
}=\partial _{\mu }A_{\nu }-\partial _{\nu }A_{\mu }$, with $A_{\mu }$ the
electromagnetic potential. In \eqref{actionmodel}, we have introduced the
gauge covariant derivative of the field with charge $q$ and its conjugate
with charge $-q$ as 
\begin{equation}
D_{\mu }\Psi =\partial _{\mu }\Psi +iqA_{\mu }\Psi \,,\qquad (D_{\mu }\Psi
)^{\star }=\partial _{\mu }\Psi ^{\star }-iqA_{\mu }\Psi ^{\star }\,,
\end{equation}%
and hereafter $\nabla _{\mu }$ denotes the covariant derivative constructed
with the Christoffel symbol. In the Abelian-Higgs model, the
self-interacting potential $V(\Psi )$ of the complex scalar field is given
by 
\begin{equation}
V(|\Psi |)\,=\,\lambda \,(\Psi \Psi ^{\star }-\nu _{0}^{2})^{2}\,,
\end{equation}%
where $\nu_0 $ is a real constant and $\lambda >0$. The field equations that
follows from varying the action \eqref{actionmodel} are \nopagebreak[3]
\begin{subequations}
\begin{eqnarray}
R_{\mu \nu }-\frac{1}{2}g_{\mu \nu }R+\Lambda g_{\mu \nu } &=&T_{\mu \nu }\,,
\label{EinsteinEq} \\
\nabla _{\mu }F^{\mu \nu } &=&J^{\nu }\,,  \label{MaxwellEq} \\
\nabla _{\mu }\nabla ^{\mu }\Psi +iq\nabla _{\mu }A^{\mu }\Psi +2iqA^{\mu
}\nabla _{\mu }\Psi -q^{2}A_{\mu }A^{\mu }\Psi -\frac{\partial }{\partial
\Psi ^{\star }}V(|\Psi |) &=&0\,.  \label{PsiEquation}
\end{eqnarray}%
\noindent The stress-energy tensor $T_{\mu \nu }$ is the sum of two
contributions 
\end{subequations}
\begin{equation}
T_{\mu \nu }\,=\,T_{\mu \nu }^{(A)}+T_{\mu \nu }^{(\Psi )}\,,
\end{equation}%
associated to the Maxwell and the scalar field, respectively, given by 
\begin{eqnarray}
T_{\mu \nu }^{(A)} &=&\frac{1}{2}\left(F_{\mu \alpha }\,{F_{\nu }}^{\alpha }-\frac{1}{4}%
g_{\mu \nu }F_{\alpha \beta }F^{\alpha \beta }\right)\,, \\
T_{\mu \nu }^{(\Psi )} &=&\frac{1}{2}\left(D_{\mu }\Psi \,(D_{\nu }\Psi )^{\star }+D_{\nu
}\Psi \,(D_{\mu }\Psi )^{\star }-g_{\mu \nu }\left( D_{\alpha }\Psi
\,(D^{\alpha }\Psi )^{\star }+V(\Psi )\right)\right) \,.
\end{eqnarray}%
In \eqref{MaxwellEq}, the particle number current is given by 
\begin{equation}
J_{\mu }\,=\,iq\left( (D_{\mu }\Psi )^{\star }\,\Psi -D_{\mu }\Psi \,\Psi
^{\star }\right) \,.
\end{equation}

In the following, we focus on two new families of independent solutions to
this model. Firstly, we construct new charged \textit{pp}-waves and (A)dS waves and
show that they are controlled by an integrable system. Then, inspired by an
extension of these solutions we will construct new charged spacetimes that
contain a two-dimensional sector whose conformal factor leads to the
Liouville equation in two dimensions. We will see that this function plays
the role of a potential and source of the Maxwell equation and the remaining
Einstein equations, respectively.

Before proceeding with the construction of the exact solutions, a few remarks are in order regarding the persistent character of the $U(1)$ currents. In \cite{wittenstrings} the deep and consequential idea of superconducting
strings was proposed. This idea (which was further generalized, for
instance, in \cite{subsequent1,subsequent2,subsequent3,subsequent4,subsequent5,subsequent6,supstr1,importance1,importance2,importance3,importance4,importance5,importance6,importance7,new10,new11,new12,new13,new14,new15,new16,new17,new18,new19}\ and
references therein) shed light on the highly non-trivial gravitational
effects of superconducting currents. These references partly motivated the
present analysis to build the simplest possible analytic example of
gravitational fields sourced by currents with the characteristics listed
here below. As far as the present analysis is concerned, the relevant features of the $%
U(1)$ persistent current \cite{wittenstrings} for our construction are:
\begin{itemize}
\item The $U(1)$ current (whose gravitational effects are under
examination) should survive even in the limit of zero gauge potential.

\item The corresponding residual current $%
J_{(0)\mu }$ (in the limit $A_{\mu }=0$)  should have the form%

\begin{equation}
J_{(0)\mu }\,=\, \Gamma \, \partial _{\mu }\Omega \ ,  \label{left3}
\end{equation}%
where $\Gamma $ is a function which cannot vanish everywhere while the
function $\Omega $ is defined only modulo $2\pi $: $\Omega \sim \Omega
+2\pi $.
\end{itemize}
As far as the function $\Gamma $ is concerned, the simplest case corresponds
to $\Gamma =cte$: in the following, we will consider configurations in
which this option is realized. While for the function $\Omega $, we will consider configurations in which the fact that $\Omega $
is defined only modulo $2\pi $ is manifest. In particular, from Eqs. (\ref%
{defin1}) and (\ref{defin3}) below it is clear that the
function $\Omega $ is defined only modulo $2\pi $ and that the current is
proportional to $\partial _{\mu }\Omega $.


\section{\textit{pp} and AdS waves}

\label{Examples} 
In order to simplify the presentation of the new solutions obtained in this
section, we separate the analysis of the $\Lambda=0$ case, from that with
non-vanishing $\Lambda$.

\subsection{\textit{pp}-waves}

The metric for a \textit{pp}-wave in Brinkmann coordinates reads 
\begin{equation}  \label{ppwave}
ds^{2} \,=\,-F\left( u,x,y\right) du^{2}-2dudv+dx^{2}+dy^{2}\,.
\end{equation}
This geometry is characterized by possessing a covariantly constant vector $%
\partial_v$, which being non-twisting, is orthogonal to the two-dimensional,
planar hypersurface spanned by the coordinates $(x,y)$. In vacuum, Einstein
equations imply that the wave profile $F(u,x,y)$ can be separated as an
arbitrary function of the coordinate $u$, times a harmonic function on $(x,y)
$. On the other hand, these spacetimes are consistent with the backreaction
produced for example by a conformal source \cite{AyonBeato:2006jf}. Metrics of the
form \eqref{ppwave} also play an important role in holography since they
emerge, for example, as supersymmetric configurations by taking a suitable
Penrose limit of the AdS$_5\times S^5$ solution of Type-IIB SUGRA
\cite{Berenstein:2002jq}.

Here, in the context of the gravitating Abelian-Higgs model %
\eqref{actionmodel}, we focus on the spontaneously broken phase, but
maintaining the phase of the scalar turned on, and we impose that both, the
gradient of the scalar $\partial _{\mu }\psi $ as well as the gauge field $%
A_{\mu }$, to be aligned with the covariantly constant vector $\partial _{v}$
that defines the \textit{pp}-wave \eqref{ppwave}. This kind of strategy to decouple
the field equations describing gauged solitons in the low energy limit of
QCD, minimally coupled to Maxwell equations, has been introduced in \cite{crystal1,gaugsk,Canfora:2018clt,lastEPJC,crystal2,crystal3,firstube}. These conditions lead to 
\begin{align}
\psi & =\rho \,e^{i\Omega (u)}\,, \quad \text{ and } \quad A=a(u,x,y)\, du\,,
\label{defin1} \\
A^{2}& =\nabla _{\mu }A^{\mu }=A\cdot \nabla \psi =0\, , \label{defin2}
\end{align}%
which have also been useful in the construction of static and rotating solutions in vector Galileon theories \cite{Cisterna:2016nwq}.

Under these circumstances, considering a constant value of $\rho=\nu_0=cte $, we
obtain an effective system of equations given by 
\begin{align}
G_{\mu \nu }& =\frac{1}{2}F_{\mu \lambda }{F^{\lambda }}_{\nu }-\frac{1}{8}%
g_{\mu \nu }F^{2}+\rho ^{2}[\nabla _{\mu }\Omega \nabla _{\nu }\Omega
+q(A_{\mu }\nabla _{\nu }\Omega +A_{\nu }\nabla _{\mu }\Omega )+q^{2}A_{\mu
}A_{\nu }]\ , \\
\nabla _{\mu }F^{\mu \nu }& =2q\rho ^{2}\nabla ^{\nu }\Omega +2q^{2}\rho
^{2}A^{\nu }\,,\label{Feff}
\end{align}%
and the Klein-Gordon equation is automatically satisfied when $\rho $ is constant.  As expected, on the spontaneously broken phase, the vector field $A_{\mu}$ acquires a mass which can be read from \eqref{Feff} leading to 
\begin{equation}\label{massA}
 m_{\textbf{A}}^2=2q^2\rho^2 \ .
 \end{equation}
Defining $\omega (u)=\partial _{u}\Omega (u)$, one can
show that the whole system for the gravitating Abelian-Higgs model, in this
sector reduces to the following two equations 
\begin{align}
\left( \frac{\partial ^{2}}{\partial x^{2}}+\frac{\partial ^{2}}{\partial
y^{2}}\right) a(u,x,y)-2q\rho ^{2}\left( qa(u,x,y)+\omega (u)\right) & =0\,,
\label{Maxwelleff} \\
\left( \frac{\partial ^{2}}{\partial x^{2}}+\frac{\partial ^{2}}{%
\partial y^{2}}\right) F(u,x,y)-\left( \frac{\partial }{\partial x%
}a(u,x,y)\right) ^{2}-\left( \frac{\partial }{\partial y}%
a(u,x,y)\right) ^{2}-& 2\rho ^{2}(qa(u,x,y)+\omega (u))^{2}=0\,.
\label{graveff}
\end{align}%
Remarkably, we have arrived to an integrable system. The equation \eqref{Maxwelleff} is a screened Poisson equation for the gauge field component $a(u,x,y)$, which can be integrated in terms of a convolution of the Green function for this operator and the source $\omega(u)$, which is the phase of the complex scalar field. Clearly, the effective mass of the vector field $m_{\textbf{A}}$ given in \eqref{massA} is responsible for the screening. Once this equation is integrated, equation \eqref{graveff} transforms into a Poisson equation for the $pp$-wave profile $F(u,x,y)$, which again, can be integrated using the corresponding Green function.

Notice that one may want to remove the phase $\omega (u)$ by a gauge
transformation $\psi \rightarrow \psi e^{-iq\xi (x^{\mu })}$ and $A_{\mu
}\rightarrow A_{\mu }+\partial _{\mu }\xi $. This can be achieved locally,
but since the phase depends on the null direction $u$, in order to remove
it, one must in general implement a large gauge transformation. Such
transformations can modify the global interpretation of the solution, thus
we prefer not to remove $\omega (u)$.


\subsection{(A)dS waves}

In this case the spacetime takes the form 
\begin{equation}
ds^{2}=\frac{\ell^{2}}{x^{2}}\left( -F\left( u,x,y\right)
du^{2}-2dudv+dx^{2}+dy^{2}\right)\, ,
\end{equation}
where we have defined $\Lambda=-3/\ell^{2}$, with $\ell$ the AdS radius. These Siklos spacetimes correspond
to a conformal transformation of the \textit{pp}-wave \eqref{ppwave}. Now, the field
equations reduce to: 
\begin{small}
\begin{align} \label{EinsAdS}
x^{4}\left( \frac{\partial^{2}}{\partial x^{2}}+\frac {\partial^{2}}{%
\partial y^{2}}\right)
a(u,x,y)-2x^{2}\ell^{2}q\rho^{2}(qa(u,x,y)+\omega(u))&=0\ , \\
\left( \frac{\partial^{2}}{\partial x^{2}}+\frac{\partial^{2}}{%
\partial y^{2}}-\frac{2}{x}\frac{\partial}{\partial x}\right) F(u,x,y)-\frac{x^2}{\ell^2}\left(\left( \frac{\partial}{\partial x}a(u,x,y)\right) ^{2}+
\left( \frac{\partial}{\partial y}a(u,x,y)\right)
^{2}\right)-2\rho^{2}(qa(&u,x,y)+\omega(u))^{2}=0\,.
\end{align}
\end{small}
In the presence of the cosmological term, the Maxwell equation \eqref{EinsAdS} is
not an autonomous equation anymore, nevertheless it can be integrated and
leads to 
\begin{equation}  \label{emfield}
a\left( u,x,y\right) =\omega\left( u\right) \left[ x^{1/2}\left(
A(y)x^{\nu}+B(y)x^{-\nu}+\left( C_{1}J_{\nu}\left( cx\right)
+C_{2}Y_{\nu}\left( cx\right) \right) \left( C_{3}e^{cy}+C_{4}e^{-cy}\right)
\right) -\frac{1}{q}\right]\ ,
\end{equation}
where $J_{\nu}$ and $Y_{\nu}$ are the Bessel functions of the first and second
kind, respectively, $A(y)$ and $B(y)$ are arbitrary linear functions of $y$
and $C_{i=1,...,4}$ and $c$ are integration constants. We have also defined 
\begin{equation}\label{nu}
\nu=\frac{1}{2}\sqrt{1+8\ell^{2}q^{2}\rho^{2}}\,.
\end{equation}
Even though we have been able to integrate the Maxwell’s equation in a closed form, for non-vanishing constants $C_i$, the metric profile $F(u,x,y)$ cannot be
integrated in a closed manner. To move forward, we therefore set $C_1=C_2=0$
as well as $A(y)=A_0$ and $B(y)=B_0$. Under these conditions the
electromagnetic field \eqref{emfield} reduces to 
\begin{equation}\label{EMads}
a\left( u,x,y\right) \,=\,\omega\left( u\right) \left[ x^{1/2}\left(
A_0x^{\nu}+B_0x^{-\nu} \right) -\frac{1}{q}\right]\ ,
\end{equation}
and the AdS-wave profile reads, 
\begin{align}
\nonumber F\left( u,x,y\right) & =\omega\left( u\right) ^{2}\left[ \left(
D_{1}e^{hx}\left( 1-hx\right) +D_{2}e^{-hx}\left( 1+hx\right) \right) \left(
D_{4}\sin\left( hy\right) +D_{5}\cos\left( hy\right) \right) \right. \\
& \left. +\,\frac{1}{8}\,\left( E_{1}+E_{2}x^{3}+\frac{1}{\ell^{2}}\left( A_{0}^{2}\frac{%
\left( 1+2\nu\right) }{\left( 3+2\nu\right) }x^{3+2\nu}+B_{0}^{2}\frac{%
\left( 2\nu-1\right) }{\left( 2\nu-3\right) }x^{3-2\nu}\right) \right) %
\right]\ .
\end{align}
Here again, $D_{1,...,5}$, $h$ and $E_{1,2}$ are integration constants. We
can see that the effect of the charge on the function $F(u,x,y)$ induces a
quite non-trivial profile.

Before finishing this section, it is interesting to notice that the expression for $\nu$ in \eqref{nu} can be written in terms of the effective mass of the vector field on the broken phase, $m_{\textbf{A}}$ given in \eqref{massA}, as
\begin{equation}
2\nu=\sqrt{1-\frac{m_{\textbf{A}}^2}{m_{\textbf{BF}}^2}}\,,
\end{equation}
where $m_{\textbf{BF}}^2=-(2\ell)^{-2}$ is the Breitenlohner-Freedman bound for a spin 1 field on AdS$_4$. Therefore, as expected, the $x$ dependence of equation \eqref{EMads} is reminiscent of that for a massive vector on AdS (see e.g. equation (25)-(26) of \cite{lYi:1998akg}).

\section{Kundt spaces in the Abelian-Higgs model}

Let us consider now consider an extension of the \textit{pp}-wave ansatz, given by 
\begin{equation}
ds^{2}\,=\,\left( f(x,y)+f_{0}v+\Lambda v^{2}\right) du^{2}-2\,dv\,du+e^{\beta h(x,y)}\,\left( dx^{2}+dy^{2}\right) \,,  \label{metricliouville}
\end{equation}%
where $u=t+w$, $v=t-w$ is a null coordinate, $x,y,w$ are Cartesian-like
coordinates, and $f_{0}$ and $\beta$ are arbitrary constants. This spacetime belongs to the Kundt family since it can be checked that the null congruence generated by $\partial_v$ is not covariantly constant, but nevertheless it has vanishing expansion, shear and twist.

The techniques developed in \cite{crystal1,gaugsk,Canfora:2018clt,lastEPJC,crystal2,crystal3} are
particularly suitable to analyze gravitating solitons whose metrics have the
form in Eq. (\ref{metricliouville}) (see the analysis in \cite{firstube}). Again, the complex scalar field adopts a harmonic dependence in $u$, and with a
constant amplitude, given by 
\begin{equation}
\Psi (x^{\mu })\,=\,\nu _{0}\,e^{i\Omega (u)}\,, \qquad \Omega (u)\,=\,u\,.
\label{defin3}
\end{equation}%
We also assume that the Maxwell field has the following form 
\begin{equation}
A=A_{u}(x,y)du\,,\qquad A_{u}(x,y)\,=\,\frac{a(x,y)-1}{q}\,.
\end{equation}%
With this ansatz, the Klein-Gordon \eqref{PsiEquation} is automatically
satisfied, while Maxwell equations reduce to 
\begin{equation}
\left( \frac{\partial^2}{\partial x^2}+\frac{\partial^2}{\partial y^2}-2\,q^{2}\nu _{0}^{2}e^{\beta h(x,y)}\right)
a(x,y)\,=\,0\,.  \label{maxwell-liouville}
\end{equation}%
The only non-trivial Einstein field equations \eqref{EinsteinEq} for this
configuration are \nopagebreak[3]
\begin{subequations}
\label{Einsteinsystem}
\begin{eqnarray}
\left( \frac{\partial^2}{\partial x^2}+\frac{\partial^2}{\partial y^2}\right)\,h(x,y) &=&-\frac{2\Lambda}{\beta}%
\,e^{\beta h(x,y)}\,,  \label{LiouvilleEq} \\
\left( \frac{\partial^2}{\partial x^2}+\frac{\partial^2}{\partial y^2}\right)\,f(x,y) &=&-\,\rho (x,y)\,,  \label{PoissonEq}
\end{eqnarray}%
\noindent with 
\end{subequations}
\begin{equation}
\rho (x,y)\,=\,2 \nu _{0}^{2}e^{\beta h(x,y)}a^{2}(x,y)+%
\frac{1 }{q^{2}}\left( \left( \frac{\partial a(x,y)}{\partial x}%
\right) ^{2}+\left( \frac{\partial a(x,y)}{\partial y}\right) ^{2}\right) \,.
\end{equation}%
 Equation \eqref{Einsteinsystem} correspond to a Liouville equation for $h(x,y)$, namely the conformal factor of the two-dimensional space spanned by the coordinates $(x,y)$ in the metric \eqref{metricliouville}, and a Poisson equation
for the function $f(x,y)$. Thus, as it happens in \cite{firstube} in the case of Einstein-Maxwell coupled to a Non-Linear Sigma Model, the present ansatz allows a useful partial decoupling of the field equations. In particular, Eq.(\ref%
{LiouvilleEq}) allows a direct integration for $h(x,y)$. This equation actually implies that the induced metric on the $u,v=$constant surfaces is of constant curvature $\Lambda$, as it occurs in vacuum \cite{Griffiths:2009dfa}. Then, once $h(x,y)$\
is known, one can solve the Maxwell equation in Eq.(\ref{maxwell-liouville})
for $a(x,y)$ since it reduces to a
Schr\"{o}dinger-like equation in which $e^{\beta h(x,y)}$ plays the role of
the potential. Eventually, once $h(x,y)$ and $a(x,y)$ are both known, one
can solve the remaining equation, Eq.(\ref{PoissonEq}) for $f(x,y)$, since
the source term $\rho(x,y)$ is explicitly known once $h(x,y)$ and $a(x,y)$ determined. This hierarchical decoupling is the key of the strategy
developed in \cite{crystal1,gaugsk,Canfora:2018clt,lastEPJC,crystal2,crystal3}. Therefore, following this logic, we start considering the general solution of \eqref{LiouvilleEq}, given by \cite{Crowdy:1997,Liouville:paper} \nopagebreak[3] 
\begin{subequations}
\label{Liouvillesolution}
\begin{eqnarray}
e^{\beta h(x,y)} &=&\frac{4}{\Lambda}\,\frac{g^{\prime }(z)\bar{g}^{\prime }(\bar{z})%
}{(g(z)\bar{g}(\bar{z})+1)^{2}}\,,\qquad \qquad \text{if}\quad \Lambda>0\,,
\label{Liouvillesolution1} \\
e^{\beta h(x,y)} &=&-\frac{4}{\Lambda}\,\frac{g^{\prime }(z)\bar{g}^{\prime }(\bar{z})}{%
(g(z)\bar{g}(\bar{z})-1)^{2}}\,,\qquad \quad \text{if}\quad \Lambda<0\,,
\end{eqnarray}%
\noindent where $g(z)$ is any meromorphic function of $z=x+iy$, with at
most simple poles, and $dg/dz\neq 0$ for all $z$ in a simply connected
domain. On the other hand, since the coordinates $x,y$ are Cartesian, the Poisson equation possesses the particular solution \cite{BookPolyanin}
\end{subequations}
\begin{equation}
f(x,y)=\frac{1}{2\pi }\,\int_{-\infty }^{\infty }\int_{-\infty }^{\infty
}\rho (\bar{x},\bar{y})\,\text{ln}\left( \frac{1}{\sqrt{(x-\bar{x})^{2}+(y-%
\bar{y})^{2}}}\right) \,d\bar{x}d\bar{y}\,.
\end{equation}%

As mentioned, the Liouville equation on $h(x,y)$, implies that the manifold spanned by the coordinates $(x,y)$ is of constant curvature $\Lambda$. Therefore, locally, there is always a change of coordinates that allows rewriting the metric \eqref{metricliouville} as
\begin{equation}
ds^{2}\,=\,\left( f(\mu,\phi)+f_{0}v+\Lambda v^{2}\right) du^{2}-2\,dv\,du+ \frac{d\mu^{2}}{1-\Lambda \mu^2}+\mu^2d\phi^{2} \,.  \label{metricliouvillertheta}
\end{equation}
In these coordinates, the equation for the electromagnetic field $a(r,\theta)$ reads

\begin{equation}\label{laeqa}
\frac{d^{2}a\left(\mu,\phi\right)  }{d\mu^{2}}+\frac{(1-2\Lambda \mu^{2}%
)}{\mu\left(  1-\Lambda\mu^{2}\right)  }\frac{da\left( \mu,\phi\right)  }%
{d\mu}-\frac{2q^{2}\nu_{0}^{2}a(\mu,\phi)}{\left(  1-\Lambda \mu^{2}\right)
}+\frac{1}{\mu^{2}\left(  1-\Lambda \mu^{2}\right)  }\frac{d^{2}a\left(\mu,\phi\right)  }{d\phi^{2}}=0\ .
\end{equation}
The general solution to this equation is
\begin{equation}
a(\mu,\phi)=\sum_m a_m\sin{(m\phi+\delta_m)}G_m(\mu) \ ,
\end{equation}
where the function $G_m(\mu)$ can be integrated in terms of Legendre functions. Here $a_m$ and $\delta_{m}$ are integration constants.

For the $\Lambda<0$ case, setting $\Lambda=-1$, the radial coordinate $\mu$ goes from $[0,\infty[$, and the solution for $G_m(\mu)$ which is non-divergent as $\mu\rightarrow\infty$ reads
\begin{equation}\label{Gmenos}
G^{\Lambda<0}_m(\mu)=\mu^{-\frac{1}{2} \left(1+2\nu\right)} \, _2F_1\left(-\frac{m}{2}+\frac{\nu}{2}+\frac{1}{4},\frac{m}{2}+\frac{\nu}{2}+\frac{1}{4},1+\nu,-\mu^{-2}\right)\,,
\end{equation}
where $_2F_1$ stands for the Gauss hypergeometric function and $\nu$ was defined in \eqref{nu}. In this case the two-dimensional surfaces at $u,v=$constant are hyperbolic spaces with origin at $\mu=0$. Even though the behavior at $\mu\rightarrow\infty$ is regular, these solutions have a singular behavior near the center $\mu=0$ of the hyperbolic space, since near such point, one can see that
\begin{equation}\label{near0}
G_m^{\Lambda<0}(\mu)=A\mu^m\left(1+\mathcal{O}(\mu)\right)+B\mu^{-m}\left(1+\mathcal{O}(\mu)\right)\,,
\end{equation}
and one can see that both constant $A,B$ are always non-vanishing. In spite of this behavior, one can check that the curvature invariants $R,R_{\alpha\beta\gamma\delta}R^{\alpha\beta\gamma\delta},R_{\alpha\beta}^{\ \ \gamma\delta}R_{\gamma\delta}^{\ \ \tau \sigma}R_{\tau\sigma}^{\ \ \alpha\beta}$
are actually constant, therefore there is no singular backreaction on the geometry.
Interestingly, the equation for $G_m$ in this case can be written as a Schr\"{o}dinger-like equation of the form
\begin{equation}
-\frac{d^2G_m(s)}{ds^2}+\frac{2q^2\nu_0^2}{\sinh^2(s)}G_m(s)=-m^2G_m(s)\ ,
\end{equation}
where we have introduced the inversion $\mu=(\sinh(s))^{-1}$ which maps the range $\mu\in(0,\infty)$ to $s\in(\infty,0)$. This is a Schr\"{o}edinger-like equation in a generalized P\"{o}schl-Teller potential, which belongs to a class of exactly solvable, shape invariant potentials \cite{FluggePractical,Cooper:1994eh}. The potential being positive, clearly implies that there cannot be solutions that are regular at both boundaries of the domain $s\in(\infty,0)$, which is consistent with the asymptotic expansion of \eqref{Gmenos} around $\mu=0$ presented in \eqref{near0}. Nevertheless, as also mentioned above, the backreaction on the geometry of this Maxwell field is regular.

When $\Lambda>0$, the range of the $\mu$-coordinate in \eqref{metricliouvillertheta} is $\mu\in\,]-1,1[\,.$ Setting $\Lambda=1$ in this case, and defining $\mu=\sin(\theta)$, leads to the following solution
\begin{equation}
G_m^{\Lambda>0}=\sin^{|m|}(\theta) \, _2F_1\left(\frac{|m|}{2}-\frac{1}{4} \sqrt{1-8 q^2 \nu_0^2}+\frac{1}{4},\frac{|m|}{2}+\frac{1}{4} \sqrt{1-8q^2 \nu_0^2}+\frac{1}{4},1+|m|,\sin^{2}(\theta)\right)\ ,
\end{equation}
which is regular at the poles located at $\theta=0$ and $\theta=\pi$.

Finally, it is also instructive to see explicitly how these cases emerge from a suitable choice of the arbitrary function $g(z)$ of the general solution of the Liouville equation in \eqref{Liouvillesolution} and \eqref{Liouvillesolution1}. For concreteness, let us focus on the case with negative cosmological constant, normalized as $\Lambda=-1$, namely the case corresponding to Eq. \eqref{Liouvillesolution1}. Choosing $g(z)=z$ and $\bar{g}(\bar{z})=\bar{z}$ in \eqref{Liouvillesolution1} leads to the following metric for the Kundt spaces 

\begin{equation}
ds^2 \, = \, (f(x,y)+f_0 v+ \Lambda v^2)du^2- 2\, dv \, du+\frac{4\,(dx^2+dy^2)}{\left(1-\left(x^2+y^2\right)\right)^2}\ ,
\end{equation}
which after the change of coordinates
\begin{equation}
x=\mu^{-1}\left(\sqrt{1+\mu^2}-1\right)\cos(\phi)\,,\qquad y=\mu^{-1}\left(\sqrt{1+\mu^2}-1\right)\sin(\phi)\ ,
\end{equation}
leads to the metric
\begin{equation}
ds^2 \, = \, (f(x,y)+f_0 v+ \Lambda v^2)du^2- 2\, dv \, du+\frac{d\mu^2}{1+\mu^2}+\mu^2d\phi^2 \ ,
\end{equation}
that we have used in \eqref{metricliouvillertheta} and \eqref{laeqa}.

\section{Conclusions}

\label{conclusions} We have constructed three new families of analytic solutions of the gravitating Abelian-Higgs model, characterized by a non-vanishing superconducting current. The first two families of solutions correspond to exact gravitational waves: \textit{pp} and (A)dS waves. In these families the null vector characterizing both the \textit{pp}-wave and the (A)dS-wave is aligned with the superconducting current. Then, we have studied a class of solutions that belong to the family of Kundt spaces, and as in vacuum, the two- dimensional geometry of the surfaces orthogonal to the superconducting currents is determined by the two-dimensional Liouville equation. Such surfaces can have either positive or negative Gaussian curvature depending on the sign of the cosmological constant.\footnote{It would be nice constructing an exhaustive classification of Kundt solutions in this model, along the lines of \cite{Ortaggio:2018zze}. } This sector possesses a remarkable property: the arbitrary analytic function characterizing the solution of the two-dimensional Liouville equation (which determines the geometry of two-dimensional surfaces transverse to the superconducting current) can be chosen in such a way that the corresponding Maxwell equations reduce consistently to a Schr\"{o}dinger-like equation in a generalized P\"{o}schl-Teller potential. Requiring suitable boundary conditions for the Maxwell field within this sector, for a negative cosmological constant, the combined effects of the gravitational and scalar interactions can confine the electromagnetic field within a bounded region of the surfaces transverse to the current itself. This result opens the interesting possibility to analyze the properties of test electromagnetic fields propagating within these families of analytic solutions of the Abelian-Higgs model using the well known properties of the P\"{o}schl-Teller potential \cite{Cooper:1994eh}. We hope to come back on this feature in the future. 

\subsection*{Acknowledgements}

We thank Eloy Ay\'{\o}n-Beato and Francisco Correa for enlightening comments on related topics. F. C., A. C. and J. O.
have been funded by Fondecyt Grants 1200022, 1210500 and 1181047, respectively. DH is partially founded by ANID grant \# 21160649.
The Centro de Estudios Cient\'{\i}ficos (CECs) is funded by the Chilean
Government through the Centers of Excellence Base Financing Program of ANID. This work is also partially funded by Proyecto de
Cooperaci\'on Internacional 2019/13231-7 FAPESP/ANID. 
\appendix

\end{document}